\pdfoutput=1
\documentclass[11pt]{article}
\usepackage[citecolor=blue]{hyperref}
\usepackage{amsmath,amsfonts,amssymb,amsthm}
\usepackage{mathrsfs}
\usepackage{verbatim}           
\usepackage{enumerate}
\usepackage[pdftex]{graphicx}
\usepackage[pdftex]{color}         
\usepackage{amscd}
\usepackage[footnotesize]{caption}

\usepackage{epstopdf}

\numberwithin{equation}{section}
\DeclareMathOperator*{\argmin}{arg\,min}

\title{Forecasting the 2013--2014 Influenza Season using Wikipedia}
\author{ Kyle S. Hickmann$^{\thanks{
             Los Alamos National Laboratory, 
             hickmank@lanl.gov}}$, 
         Geoffrey Fairchild$^{*}$, 
         Reid Priedhorsky$^{*}$,
         Nicholas Generous$^{*}$ \\
         James M. Hyman$^{\thanks{
           Tulane University, Mathematics Department}}$, 
         Alina Deshpande$^{*}$, 
         Sara Y. Del Valle$^{*}$
}

\date{}
\begin{document}
\maketitle

\begin{abstract}
  Infectious diseases are one of the leading causes of morbidity and mortality
  around the world; thus, forecasting their impact is crucial for planning an
  effective response strategy. According to the Centers for Disease Control
  and Prevention (CDC), seasonal influenza affects between 5\% to 20\% of the
  U.S. population and causes major economic impacts resulting from
  hospitalization and absenteeism. Understanding influenza dynamics and
  forecasting its impact is fundamental for developing prevention and
  mitigation strategies.

  We combine modern data assimilation methods with Wikipedia access logs and
  CDC influenza-like illness (ILI) reports to create a weekly forecast for
  seasonal influenza. The methods are applied to the 2013--2014 influenza
  season but are sufficiently general to forecast any disease outbreak, given
  incidence or case count data. We adjust the initialization and
  parametrization of a disease model and show that this allows us to determine
  systematic model bias.  In addition, we provide a way to determine where the
  model diverges from observation and evaluate forecast accuracy.

  Wikipedia article access logs are shown to be highly correlated with
  historical ILI records and allow for accurate prediction of ILI data several
  weeks before it becomes available. The results show that prior to the peak
  of the flu season, our forecasting method projected the actual outcome with
  a high probability. However, since our model does not account for
  re-infection or multiple strains of influenza, the tail of the epidemic is
  not predicted well after the peak of flu season has past.
\end{abstract}


\medskip


\medskip


\section*{Author Summary}

We use modern methods for injecting current data into epidemiological
models in order to offer a probabilistic evaluation of the future
influenza state in the U.S. population. This type of disease
forecasting is still in its infancy, but as these methods become more
developed it will allow for increasingly robust control measures to
react to and prevent large disease outbreaks. While weather
forecasting has steadily improved over the last half century and
become ubiquitous in modern life, there is surprisingly little work on
infectious disease forecasting. Although there has been a great deal
of work in modeling disease dynamics, these have seldom been used to
generate a probabilistic description of expected future dynamics,
given current public health data. Moreover, the mechanism to update
expected disease outcomes as new data becomes available is just
beginning to receive attention from the public health community. Using
CDC \emph{influenza like illness} reports and digital monitoring
sources, such as observations of Wikipedia access logs, we are now at
a point where forecasting for the influenza season can be brought
inline with our advancements in weather prediction.

\section{Introduction}

Despite preventative efforts and educational activities for seasonal
influenza, thousands of people are affected every year and many die,
resulting in a significant public health and economic burden for the
U.S. population
\cite{_national_2012,fraser_pandemic_2009,germann_mitigation_2006}. The
CDC monitors influenza burden by collecting information from volunteer
public health departments at the state and local level
\cite{Burkhead2000}. Data are then used for planning and mitigation
activities based on what is believed to be the current state of
influenza throughout the U.S. \cite{_overview_2012,hopkins2014}. These
rough estimates can lead to significant over- or under-preparation for
any given flu season since little modeling of influenza dynamics is
used to extrapolate from the current state of influenza to the state
of influenza for the remaining season.

The capability for real-time forecasting of events, such as influenza
dynamics, with quantified uncertainty, has been crucial for numerous
major advances across the spectrum of science
\cite{kalnay_atmospheric_2003,evensen_data_2009,creal_survey_2012}. However,
this capability is still in its infancy in the field of public
health. For more complete literature reviews on the field we refer the
reader to \cite{chretien2014influenza,nsoesie2014systematic}. Briefly,
we present the literature on epidemic forecasting influencing this
work, all of which rely on a Bayesian viewpoint to adjust an
underlying disease model given incoming observations
\cite{bettencourt2007}. First, disease forecasting methods that use
data to parameterize an underlying causal model of disease can use
either sequential Monte Carlo type methods
\cite{bettencourt2008,nsoesie2013simulation,yang2014simple,ong2010real,skvortsov_monitoring_2012,jegat_early_2008,balcan2009seasonal}
or ensemble methods
\cite{shaman_forecasting_2012,shaman2013real,cobb2014bayesian,mandel_data_2010}. Some
work has been done on comparing the two methods
\cite{yang2014comparison,sheinson2014comparison}. There are also
several works of more statistical nature
\cite{nsoesie2011prediction,nsoesie2014dirichlet,nsoesie2013forecasting,chakrabortyforecasting2014,safta2011real,breto2009time},
one that relies on a pure Kalman filter \cite{cazelles1997using}, and
one \cite{rhodes_variational_2009} that uses variational assimilation
methods. Of these works, the majority tune a differential equation
based compartmental disease model
\cite{sheinson2014comparison,jegat_early_2008,rhodes_variational_2009,skvortsov_monitoring_2012,yang2014comparison,shaman2013real,yang2014simple,shaman_forecasting_2012,bettencourt2008}. However,
some forecasts have been formed using agent based simulations
\cite{balcan2009seasonal,nsoesie2013simulation,chakrabortyforecasting2014}
or spatial models \cite{mandel_data_2010,cobb2014bayesian}.

Each of these examples rely on defining a prior distribution for the
parameterization and initialization of the underlying model. However,
the methods to arrive at this prior are usually \emph{ad hoc} and
based on beliefs about the ranges for the parameters. It is therefore
difficult to see how these methods may be applied to a general disease
model given historic observations in the presence of model error. In
this work, we outline a method for defining a prior
parameterization/initialization that can be generalized to any data
set pertaining to disease spread and disease model.

With the exception of reference \cite{rhodes_variational_2009}, the
aforementioned methods only use the most recent observation to update
the epidemic model. This can lead to problems in determination of the
underlying model parameters since the dynamic trends of the data are
not considered during the model adjustment. We use an \emph{ensemble
  Kalman smoother} approach that is more sensitive to the underlying
dynamics of the data timeseries.

Another difficulty during data assimilation is that, in the adjustment
of the current model state, conditions such as the population in each
epidemic category summing to the total population are often
disrupted. Moreover, since the model state is adjusted each time an
observation is made, the forecast epidemic curve may not represent any
single realization of the epidemic model. This makes it difficult to
judge systematic model error and thus identify specific areas to where
the model may be improved. To remedy this difficulty in our
assimilation scheme, we only adjust the model's parameterization and
initialization. Therefore, each of our forecasts represent a
realization of the model.

In November of 2013, the CDC launched the \emph{Predict the Influenza
  Season Challenge} competition to evaluate the growing capabilities
in disease forecasting models that use digital surveillance data
\cite{cdc_challenge2013}. The competition asked entrants to forecast
the timing, peak, and disease incidence of the 2013-2014 influenza
season using Twitter or other Internet data to supplement ILI weekly
reported data.  The work described in this paper was conducted as an
entry for this competition.

The significant obstacles to accurately predicting the spread of
diseases include:
\begin{itemize}
\item Data pertaining to disease dynamics is difficult to obtain in
  real time for many diseases.  When data is available, such as ILI
  data, it can be difficult to quantify its relation to the actual
  number of disease (e.g. influenza) cases.  Also, when using recent
  data, the underreporting and reporting delays lead to additional
  uncertainties and a constantly updated database.
\item Data assimilation algorithms must be developed for dynamically
  incorporating the incidence data into mathematical transmission
  models.  Although these methodologies have been developed for
  atmospheric sciences (i.e., weather prediction) models, the
  epidemiology community has only recently begun apply these
  approaches to transmission models.
\end{itemize}

Reliable forecasts of influenza dynamics in the U.S. cannot be
obtained without consistently updated public health observations
pertaining to flu \cite{donnelly_epidemiological_2003}. It is
necessary to have a historical record of these observations in order
to asses the relation between the forecasting model and the data
source. We used the well established public health observation system
for influenza in the U.S., the \emph{influenza-like illness} (ILI)
network \cite{_overview_2012,ong2010real}. This network has been in
place for over a decade and reports the percentage of patients in the
U.S. seen for ILI each week. ILI data provide an informative depiction
of the state of influenza in the U.S. each week. However, it
represents a limited syndromic observation of people who seek medical
attention each week \cite{hopkins2014}. This is different than what
our disease model represents, which is the dynamics of the proportion
of the population currently infected. A further drawback is related to
the bureaucratic hierarchy of the ILI system; there is a 1--2 week lag
present in data availability.

The addition of other data sources that provide different estimates
for influenza incidence can add to the robustness of this data stream
\cite{nick2014,mciver2014wikipedia,ginsberg_detecting_2008,priedhorsky_inferring_2014,lampos_tracking_2010}. In
this work, we use Wikipedia access logs for articles highly correlated
with influenza prevalence, as measured by ILI, to improve our
knowledge of the current influenza incidence in the U.S. The rationale
for using the Wikipedia access logs was thoroughly explored in
reference \cite{nick2014}.

Influenza forecasting must provide two things in order to inform
public health policy: 1) the expected future influenza dynamics and 2)
the likelihood of observing dynamics deviating from this
expectation. These two properties must be informed by both the
inherent model dynamics and current observations of flu.  Fortunately
for epidemiologists, the problem of turning a deterministic
mathematical model into a probabilistic forecast using observed data
has a long history in the areas of climatology, meteorology, and
oceanography
\cite{kalnay_atmospheric_2003,evensen_data_2009,creal_survey_2012}. We
demonstrate that one of these techniques, the \emph{ensemble Kalman
  smoother}, can be used to iteratively update a distribution of the
influenza model's initial conditions and parameterizations. An
advantage of this particular technique is that it retains information
about when the model dynamics systematically diverge from the dynamics
of observations. Therefore, it is clear where to focus efforts to
improve the forecasting power of the model.

After outlining the general forecast methodology, we describe the
details of our data sources and model, the technique used to estimate
a \emph{prior} forecast, our data assimilation technique, and our
measure of forecast accuracy. We then present an application of our
methods to forecasting the 2013--2014 influenza season in the results
section, and conclude with a summary of our approach and suggestions
for future improvements.

\section{Methods}

\subsection*{Data and Model}

Our goal is to forecast future U.S. influenza dynamics using CDC ILI
data and Wikipedia access log data. On one hand, the ILI data
represent the ground truth that we attempt to forecast and use to
train our models. The Wikipedia data, on the other hand, has the
potential to provide information about the current state of influenza.

The CDC's ILI data represent the collection of outpatient data from
over 3,000 hospitals and doctors' offices across the U.S. Each week,
these locations report the total number of patient visits and the
number of those visits that were seen for ILI defined as fever
(temperature $\ge 100^{\circ}$F) and a cough or sore throat without a
known cause other than influenza. Since 2003, these data have been
collected weekly, year-round. Clearly, these data are related to the
proportion of the U.S. population infected with influenza. However, it
does not contain information about the proportion of the population
that does not seek treatment and the proportion of people seeking
treatment who have flu symptoms but not the flu virus.

ILI outpatient data make up one portion of the CDC's influenza
surveillance capabilities. The outpatient data described above are
also divided into 10 Health and Human Services (HHS) regions. In
collaboration with the World Health Organization (WHO), U.S. influenza
vaccination coverage and virological strain data are also collected,
we only use the total U.S. ILI data. For a complete overview of the
CDC's influenza surveillance programs we refer the reader to
\cite{_overview_2012,ong2010real}.

As mentioned above, there is a 1--2 week delay between a patient
seeing a doctor and the case appearing in the ILI database. Therefore,
there is a need for the use of, more real time, digital surveillance
data that can complement ILI. Wikipedia provides summary article
access logs to anyone who wishes to use them. These summaries contain,
for each hour from December 9, 2007 to present (and updated in real
time), a compressed text file listing the number of requests served
for every article in every language, for articles with at least one
request. We aggregate these hourly requests into weekly access counts
and normalize the total number of accesses per article using the total
requests for all articles across the entire English Wikipedia in each
week. This data source has been studied extensively in
\cite{nick2014,mciver2014wikipedia} and we refer the reader to these
sources for a more complete description of this data source.

Five articles from the English language edition of Wikipedia were
selected for estimation of present national ILI using the methods
outlined in \cite{nick2014}. These articles were \emph{Human Flu,
  Influenza, Influenza A virus, Influenza B virus,} and
\emph{Oseltamivir}. The weekly article request data for each article
can then be written as the independent variables $x_1, x_2, \dots,
x_5$. Current ILI data are estimated using a linear regression from
these variables. We combine the article request data with the previous
week's ILI data, which we'll denote by $ILI_{-1}$, and a constant
offset term. This forms our regression vector $X = (1, x_1, x_2,
\dots, x_5, ILI_{-1})$. Our linear model used to estimate the current
week's ILI data is then given by
\begin{equation}\label{ILI_linear_model}
  ILI_0 = b \cdot X = b_0 + b_1 x_1 + b_2 x_2 + \dots + b_5 x_5 + b_6 ILI_{-1}.
\end{equation}
The regression coefficients $b = (b_0, b_1, b_2, \dots, b_6)$ were
then determined from historical ILI data and Wikipedia data.

\subsubsection*{Model description}

We only model the U.S. ILI data during the part of the year which we
designate as the \emph{influenza season}. By examining the ILI data
for the entire U.S. from the 2002--2003 season to the 2012--2013
season, it was found that influenza incidence does not start to
noticeably increase until at least epidemiological week 32
(mid-August). Moreover, once the influenza peak has past, the
incidence decreases by at least epidemiological week 20 (mid-May). The
exception to this range is the the 2009 H1N1 pandemic, which emerged
in the late 2008--2009 season causing this season to be prolonged and
an early start in 2009--2010 season. With the flu season defined to be
between epidemiological week 32 and epidemiological week 20, even the
2009 H1N1 emergence is mostly accounted for. This allows us to avoid
modeling influenza prevalence during the dormant summer months.

The U.S. ILI data between mid-August and mid-May are then modeled
using a Susceptible-Exposed-Infected-Recovered (SEIR) differential
equation model
\cite{anderson_infectious_1991,hethcote_mathematics_2000,ross_prevention_1910}. The
standard SEIR model is then modified to allow seasonal variation in
the transmission rate \cite{hyman2003modeling} and to account for
heterogeneity in the contact structure
\cite{del_valle_mixing_2007,stroud_semi-empirical_2006,stroud_spatial_2007}. We
will refer to the model as the \emph{seasonal $S^{\nu}\!EIR$}
model. This model does not account for several factors that could
possibly be important for forecasting influenza dynamics such as
spatial disease spread, behavior change due to disease, multiple viral
strains, vaccination rates, or more detailed contact structure
\cite{alfaro2013deterministic,del_valle_modeling_2013,bajardi2011human,lee_computer_2010}.

In our model, the U.S. population is divided into epidemiological
categories for each time $t > 0$ as follows: the proportion
\emph{susceptible} to flu $S(t)$, the proportion \emph{exposed} (and
noninfectious, nonsymptomatic) $E(t)$, the proportion \emph{infectious
  and symptomatic} $I(t)$, and the proportion \emph{recovered and
  immune} $R(t)$. Since there is usually a single dominant strain each
flu season, we assume that recovered individuals are then immune to
the disease for the remainder of the season. In practice, this
assumption is not entirely accurate since an individual that
contracted and recovered from one strain can get infected from a
different strain in the same season \cite{alfaro2013deterministic}.

\begin{figure}[!h]
  \includegraphics[scale=0.5]{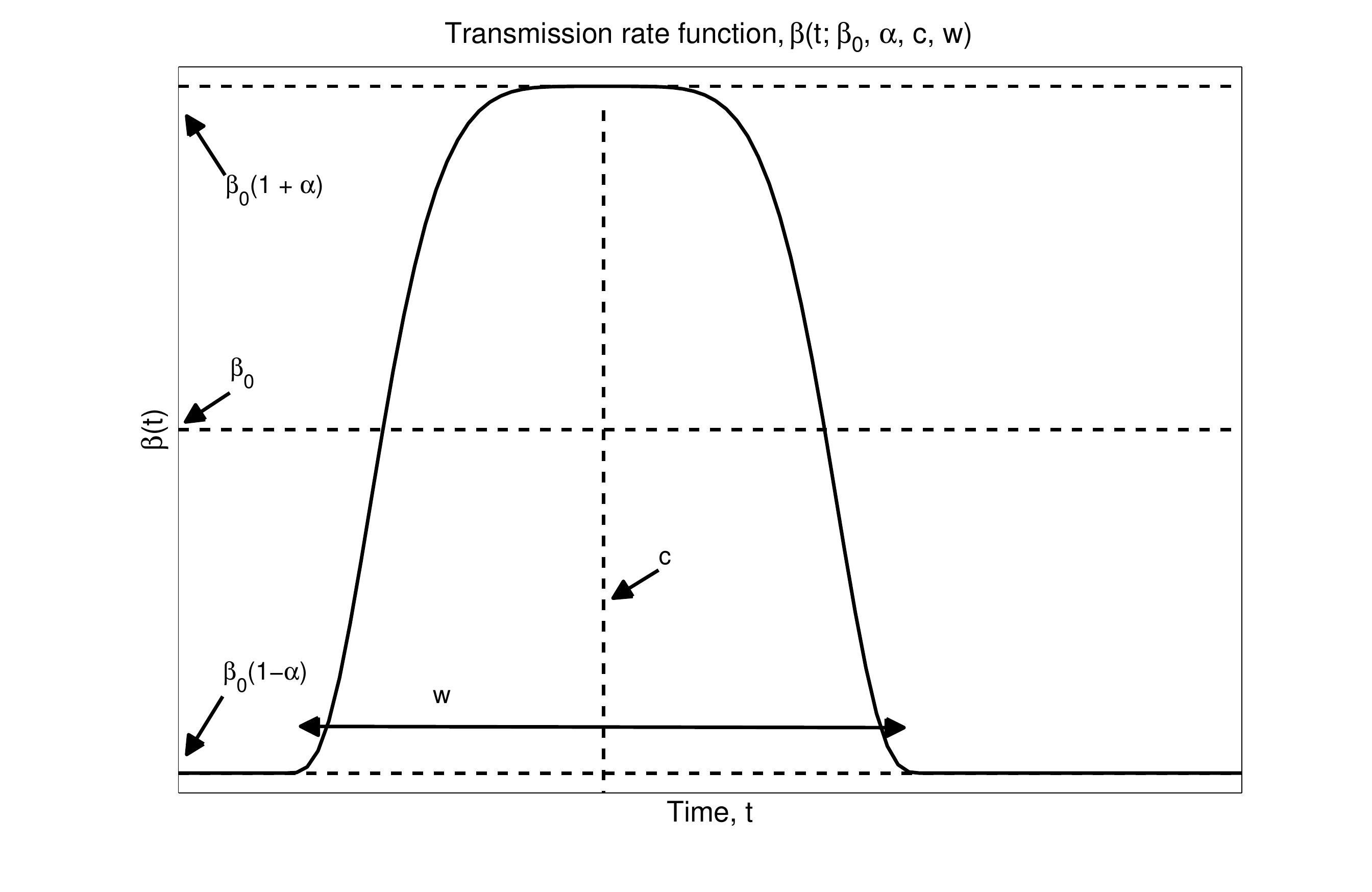}
  \caption{{\bf The transmission rate function $\beta(t; \beta_0,
      \alpha, c, w)$.}  The transmission function is chosen to be a
    smooth, five times differentiable, bump function ranging between
    $\beta_0 (1 + \alpha)$ at the peak of flu transmission and
    $\beta_0 (1 - \alpha)$ at the low point. This is done to account
    for seasonality in our model. The parameters $c$ and $w$ control
    the center (peak of elevated flu transmission) and width (duration
    of elevated flu transmission).}
  \label{fig:beta}
\end{figure}

The seasonal $S^{\nu}\!EIR$ model is defined by the following system
of ordinary differential equations:
\begin{align}\label{SEIRplus}
  \frac{dS}{dt} &= - \beta(t; \beta_0, \alpha, c, w) I S^{\nu} \\ \nonumber
  \frac{dE}{dt} &= \beta(t; \beta_0, \alpha, c, w) I S^{\nu} - \theta E \\ \nonumber
  \frac{dI}{dt} &= \theta E  - \gamma I \\ \nonumber
  \frac{dR}{dt} &= \gamma I \\ \nonumber
  S(0) &= S_0 \qquad E(0) = E_0 \qquad I(0) = I_0 \qquad R(0) = 1 - (S_0 + E_0 + I_0).
\end{align}
$S(t)$, $E(t)$, $I(t)$, and $R(t)$ are the proportions of the
U.S. population at time $t > 0$ defined above. Individuals transition
from exposed to infectious with constant incubation rate $\theta$, and
they recover at constant rate $\gamma$. The transmission coefficient,
$\beta(t; \beta_0, \alpha, c, w)$ is allowed to vary over the course
of the flu season. The specific variation is controlled by the
parameters $(\beta_0, \alpha, c, w)$, as shown in
Figure~\ref{fig:beta}. To model some aspects of heterogeneity in the
influenza contact network, we use a power-law scaling, $\nu$, on the
susceptible proportion of the population in the term
$S^{\nu}$. Including this factor has been demonstrated to be an
effective approach for this simple model to better fit large-scale
detailed agent based models with a heterogeneous contact network
\cite{stroud_semi-empirical_2006}.

\subsection*{Prior distribution estimation}

We start by specifying a distribution of model parameterizations that
we will consider before any observations from the 2013--2014 season
are available. This \emph{prior} distribution specifies what we think
is possible to observe in the new influenza season. Therefore, it is
based on the previously observed ILI data and is broad enough to
assign a high likelihood to any of the past influenza seasons. Though
our method of specifying a prior is reasonable enough to meet this
criterion, it does not rely on more rigorous approaches, this will be
left to future work.

Let us assume that we have ILI observations from $M$ different
influenza seasons. The observations for each season are made at
regular intervals, $\Delta t = 1 \textrm{ week}$, from the end of
epidemiological week 32 to the end of epidemiological week 20 of the following
year. With the seasons indexed by $i$ we denote these data by
\begin{equation}\label{data_discretization}
  d^i_{1:K} = (d^i_{\Delta t}, d^i_{2 \Delta t}, \dots, d^i_{K \Delta t})^T,
\end{equation}
with $K$ being the number of weeks in each season and $i = 1, 2,
\dots, M$.

A solution of our model is determined by the parameterization vector
\begin{equation}\label{SEIRparam}
\mathbf{p} = (S_0, E_0, I_0, \beta_0, \alpha, c, w, \theta, \gamma)^T,
\end{equation}
and each choice of $\mathbf{p}$ yields a discretely sampled solution vector
\begin{equation}\label{SEIR_discretization}
\Psi_{1:K} = (\psi^T_{\Delta t}, \psi^T_{2 \Delta t}, \dots, \psi^T_{K \Delta t}, \mathbf{p}^T)^T.
\end{equation}
The state of our $S^{\nu}\!EIR$ model is denoted by 
\begin{equation}\label{SEIRvect}
\psi_t = (S(t), E(t), I(t))^T.
\end{equation}
The link between our epidemiological model and the data is obtained
from the infected proportion at the discrete time
points. Specifically, it is the model-to-data map defined by
\begin{equation}\label{SEIRpriormodel2dataMap}
M_{\mathbf{p}}[\Psi_{1:K}] = (100 \cdot I(\Delta t), 100 \cdot I(2 \Delta t), \dots, 100 \cdot I(K \Delta t))^T,
\end{equation}
with the multiplication changing the proportion into a percentage,
which is what ILI is measured in. Our goal now is to determine a prior
distribution for $\mathbf{p}$, $\pi_0(\mathbf{p})$, so that samples
drawn from the prior, make $M_p[\Psi_{1:K}]$ close to at least one
prior season's data set, $d^i_{1:K}$.

For each season's data, $i = 1, 2, \dots, M$, we can determine an allowable
$\mathbf{p}^i$ by approximately solving the non-linear optimization problem
\begin{equation}\label{E:general_discrepancy}
  \mathbf{p}^i  = \argmin_{\mathbf{p}} \| d^i_{1:K} - M_{\mathbf{p}}[\Psi_{1:K}] \|^2,
\end{equation}
where $\| \cdot \|$ denotes the root sum of squares discrepancy over
the discrete time points. The approximate solution to
(\ref{E:general_discrepancy}) is reached by applying a stochastic
optimization algorithm \cite{Geem2001,Geem2006a} and this process is
repeated $L$ times for each season. Variation in these optimal
$\mathbf{p}$ for a single season are considered to represent variation
that we should allow in the prior distribution of our model. This
process then yields $M \cdot L$ approximate solutions
$\mathbf{p}^i_l$, which are then treated as samples from a prior
distribution for the model's parameterization.

A log-normal distribution, fit to these samples, is chosen for
$\pi_0(\mathbf{p})$.  We have chosen a log-normal distribution for
$\pi_0(\mathbf{p})$ since physically all terms in $\mathbf{p}$ must be
positive and the relation of the log-normal to a Gaussian distribution
makes it a convenient choice when implementing our ensemble Kalman
filtering method.

\subsection*{Data assimilation}

An iterative data assimilation process is implemented to continually
adjust the parameters and initial state of the seasonal $S^{\nu}\!EIR$
model, which incorporates new ILI and Wikipedia observations. The
model can then be propagated through the end of flu season to create
an informed forecast. Data assimilation has been successful in a
diverse array of fields, from
meteorology~\cite{kalnay_atmospheric_2003,evensen_data_2009} to
economics~\cite{creal_survey_2012} but has only recently begun to be
applied to disease spread
\cite{jegat_early_2008,mandel_data_2010,skvortsov_monitoring_2012,shaman_forecasting_2012,sheinson2014comparison,yang2014simple,shaman2013real,yang2014comparison}.

We use an ensemble Kalman smoother (enKS)
\cite{evensen_data_2009,evensen2000ensemble}, with propagation always
performed from the start of the influenza season, to assimilate the
ILI/Wikipedia data into the transmission model. Our implementation of
the enKS directly adjusts only the parameterization of our
system. However, the adjustment is determined using information about
the model's dynamics throughout the season.

For a more precise description of the ensemble Kalman smoother
implemented for this work we refer the reader to
\cite{evensen_data_2009}. The main idea is to view the time series of
our epidemiological model together with its parameters and the
ILI/Wikipedia data as a large Gaussian random vector. We can then use
standard formulas to condition our $S^{\nu}\!EIR$ time series and
parameters on the ILI/Wikipedia observations. Draws from this
conditional Gaussian give an updated parameterization of the system
from which we can re-propagate to form an updated forecast.

The enKS is similar to a standard ensemble Kalman filter except that
instead of just using the most recent data to inform the forecast it
uses a number of the most recent observations.  The three most current
observations, including Wikipedia observations, are used to inform our
forecasts. The advantages of using the enKS is that more of the
current trends/dynamics of the observations are used in each
assimilation step. This helps in estimating the underlying
parameterizations of the system by propagating the observation's
information backward into the model ensemble's history
\cite{evensen_data_2009,evensen2000ensemble}.

For each week in the simulation, we receive the ILI data and a
Wikipedia estimate of the ILI data. These data become available at
regular time intervals of $\Delta t = 1 \textrm{ week}$ and we denote
the data corresponding to the first $K$ weeks by $d_{1:K}$ as in
(\ref{data_discretization}). Note that now the index $K$ corresponds
to the most current week instead of the last week in the
season. During the data assimilation step, $d_{1:K}$ is compared with
simulations of our $S^{\nu}\!EIR$ model and its parameterization,
sampled at weekly intervals, denoted by $\Psi_{1:K}$ as in
(\ref{SEIR_discretization}) above.

The link between our epidemiological model and the data is again
obtained from the simulated infected proportion at the time the most
recent data are collected, $K \Delta t$. This is similar to
(\ref{SEIRpriormodel2dataMap}) except that we only use the infected
proportions corresponding to recent data. Specifically, the
\emph{model-to-data map} is
\begin{equation}\label{SEIRmodel2dataMap}
M[\Psi_{1:K}] = (100 \cdot I((K-2) \Delta t), 100 \cdot I((K-1) \Delta t), 100 \cdot I(K \Delta t))^T.
\end{equation}
Using this model-to-data map implies that we are only attempting to
model and forecast the dynamics of ILI as opposed to the actual
proportion of the U.S. population infected with influenza. The last
three sampled values of the infected proportion are used,
corresponding to the Wikipedia estimated ILI, the most current ILI,
and the previous week's ILI observations.

In the ensemble Kalman filtering framework, the simulation and data,
$(\Psi^T_{1:K}, d^T_{(K-2):K})^T$, are assumed to be jointly Gaussian
distributed. Therefore, the conditional random vector $\Psi_{1:K} |
d_{(K-2):K}$ is also Gaussian, which we can sample from. We only
sample the marginal distribution, which is also Gaussian, of our
$S^{\nu}\!EIR$ parameterization, $\mathbf{p} | d_{(K-2):K}$. Samples
of $\mathbf{p} | d_{(K-2):K}$ are then used to re-propagate our
$S^{\nu}\!EIR$ model from an adjusted initial state to form an updated
forecast. When new data are collected on the
$(K+1)$\textsuperscript{th} week, the process is repeated.

The remaining details of the enKS implementation deal with the choice
of the mean and covariance structure of the joint Gaussian
distribution for $(\Psi^T_{1:K}, d^T_{(K-2):K})^T$. Our implementation
followed Evensen's explanation \cite{evensen_data_2009}. In short, the
mean is determined by sampling our $S^{\nu}\!EIR$ model at different
parameterizations, while the covariance structure is determined by
assumptions on the observational error for ILI, the Wikipedia
estimate, and our epidemiological model.

\subsection*{Evaluating forecast accuracy}

To evaluate the accuracy of a forecast, we compare the distribution
determined by the ensemble with the actual observed disease data. We
can, of course, only perform this evaluation retrospectively since we
require data to evaluate our forecasts against. Since the enKS method
assumes that the forecast distributions are Gaussian, we can evaluate
the forecast's precision by scaling the distance of our forecast mean
from the observation using the ensemble covariance. Such a distance
has been widely used in statistics and is commonly referred to as the
\emph{Mahalanobis distance} (M-distance)
\cite{mahalanobis1936generalized}. The M-distance gives a description
of the quality of the forecast that accounts for both precision in the
mean prediction and precision in the dispersion about the mean. Other
methods of evaluating forecast accuracy such as the root mean square
error only consider how close the mean of the forecast is to the
observations. Thus, a distribution with a great deal of uncertainty,
or dispersion, can have a small root mean square error compared to
observations.

A forecast is made up of an analysis ensemble of parameterizations $\{
\mathbf{p}^i_K \}_{i=1}^N$, $K$ is the index corresponding to the most
recently assimilated observation and $N$ is the size of the
ensemble. Each $\mathbf{p}^i_K$ is drawn from a Gaussian distribution
conditioned on the most recent observations as described above. We can
form a forecast of ILI data for the entire season by propagating the
$\mathbf{p}^i_K$ through our $S^{\nu}\!EIR$ model. We will denote the
discretely sampled time series of these realizations by
$\psi^i_K$. The M-distance will then be evaluated using the ensemble
of forecast observations, $\{ M_f[\psi^i_K] \}_{i=1}^N$, corresponding
to the infected proportion time series, after the time index $K$, with
each $\psi^i_K$ scaled to a percentage.

The M-distance is then calculated from the $M_f[\psi^i_K]$ using their
sample mean and covariance denoted $\mu_{\textrm{obs}}$ and
$C_{\textrm{obs}}$, respectively. Letting $\tilde{d}_K$ correspond to
un-assimilated observations (i.e. observations with time indices
greater than $K$) the M-distance we evaluate is
\begin{equation}\label{mahalanobis_dist}
\rho(\tilde{d}_K, \{ M_f[\psi^i_K] \}) = \sqrt{(\tilde{d}_K - \mu_{\textrm{obs}})^T C^{-1}_{\textrm{obs}} (\tilde{d}_K - \mu_{\textrm{obs}})}.
\end{equation}

In order to judge the quality of our forecasting methods and
ultimately to justify the complexity of our data assimilation
procedure, we generate a simplistic \emph{straw man} model for
comparison. We first collect all historical time series of disease
outbreaks and then determine a correspondence time between each of the
time points for each of the outbreak data sets. This gives a common
time frame for each of the historical data sets. Then, at each of
these common time points, the average and standard deviation of the
historical observations can be computed. Thus, the straw man forecast
consists of a normal distribution at each corresponding time point in
the forecast with an averaged mean and standard deviation. We can then
evaluate the straw man's accuracy using the metric given in
equation~(\ref{mahalanobis_dist}). Given the simple construction of
the straw man forecast, this provides a good baseline to necessarily
beat, in terms of smaller M-distance, for any compartmental data
assimilation-derived forecast.

Besides using a strictly quantitative measure of forecast accuracy, we
also suggest computing more qualitative measures of accuracy. With the
ensemble forecast, the samples $\psi^i_K$ can be used to estimate
quantiles of the forecast distribution such as the standard
\emph{5-number summary} of the distribution given by the 5\%, 25\%,
50\%, 75\%, and 95\% quantiles for the seasonal $S^{\nu}EIR$
realizations. Moreover, if we are interested in the forecast of some
other quantity of interest derived from the time series of
observations, such as the epidemic's peak time, peak level, duration,
or start time, we may also derive 5-number summaries for these
quantities by computing the appropriate quantity of
interest. Analyzing where the actual observations fall compared to the
5-number summary provides a qualitative way to understand the accuracy
of the forecast.

\section{Results}

We present the results of our prior estimation techniques and our
evaluation of forecast accuracy for the 2013--2014 influenza
season. Again, for the reasons mentioned above, the influenza season
is defined to be between the 32\textsuperscript{nd} and
20\textsuperscript{th} epidemiological weeks for successive years. Our
forecasts are only valid during this time period.

\subsection*{Prior forecast}

\begin{figure}[!h]
  \begin{center}
  \includegraphics[width=\textwidth]{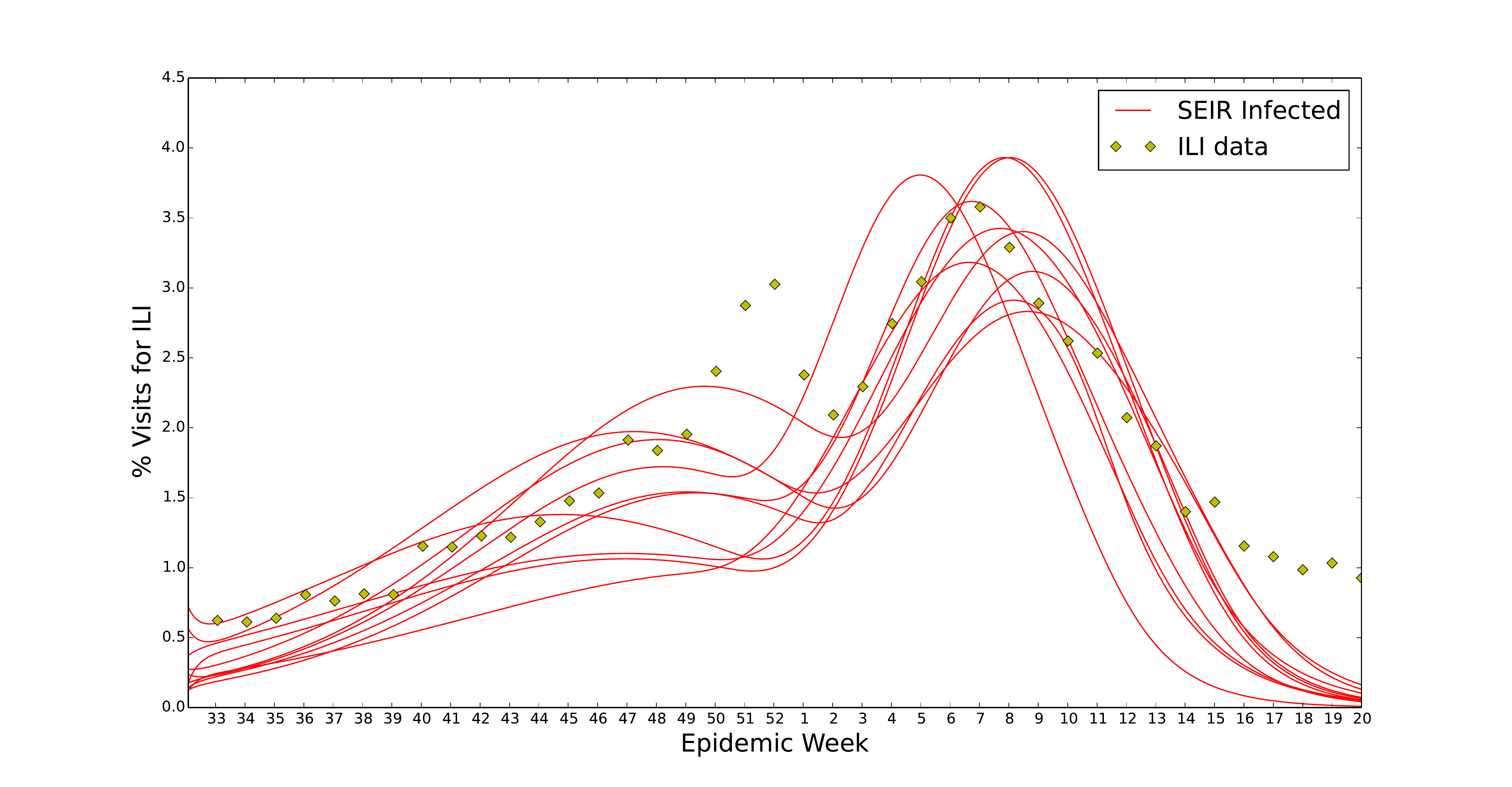}
  \caption{{\bf Seasonal $S^{\nu}\!EIR$ fit to 2006--2007 U.S. ILI
      data.} Ten seasonal heterogeneous $S^{\nu}\!EIR$ model
    parameterizations for the U.S. ILI 2006--2007 data. These are
    approximate solutions to (\ref{E:general_discrepancy}). For each
    of the influenza seasons, from 2003--2004 through 2012--2013, fits
    similar to the above were generated. These parameterizations
    formed the basis for our prior, $\pi_0(\mathbf{p})$. This is a
    good example of the seasonal $S^{\nu}\!EIR$ model's two areas of
    systematic divergence. In the weeks 50--1 there is a first peak
    that the model does not catch. During the tail weeks 15--20 our
    $S^{\nu}\!EIR$ model tapers too quickly.}
  \label{fig:US_06_07_SEIRfit}  
\end{center}
\end{figure}

Historical ILI data from the 2003--2004 U.S. influenza season through
the 2012--2013 influenza season was used to generate our prior
distribution of the seasonal $S^{\nu}\!EIR$ model's
parameterization. This was done following the methods described
above. An example fit using a stochastic optimization algorithm to
find 10 approximate solutions to (\ref{E:general_discrepancy}) for the
2006--2007 ILI data is shown in Figure~\ref{fig:US_06_07_SEIRfit}.
Two things can be noticed from this fit of our epidemiological
model. First, there is often a small early peak in the ILI data before
the primary peak and our model does a poor job of capturing
this. Second, the ILI data usually remains elevated longer than our
model's realizations can support. Both of these areas point to
systematic divergence of the model from data.

\begin{figure}
  $\begin{array}{cc}
  \includegraphics[scale=0.2]{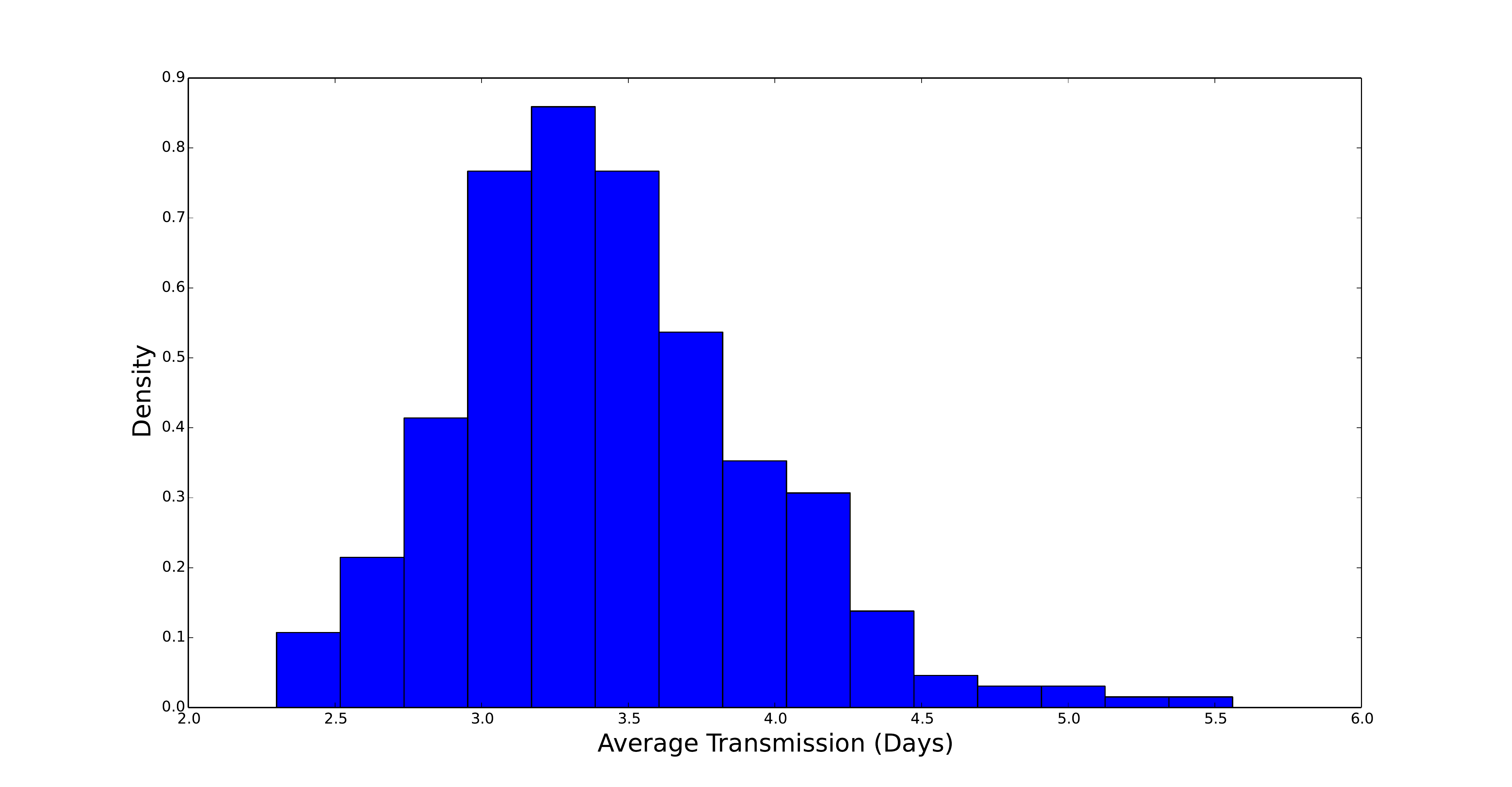} &
  \includegraphics[scale=0.2]{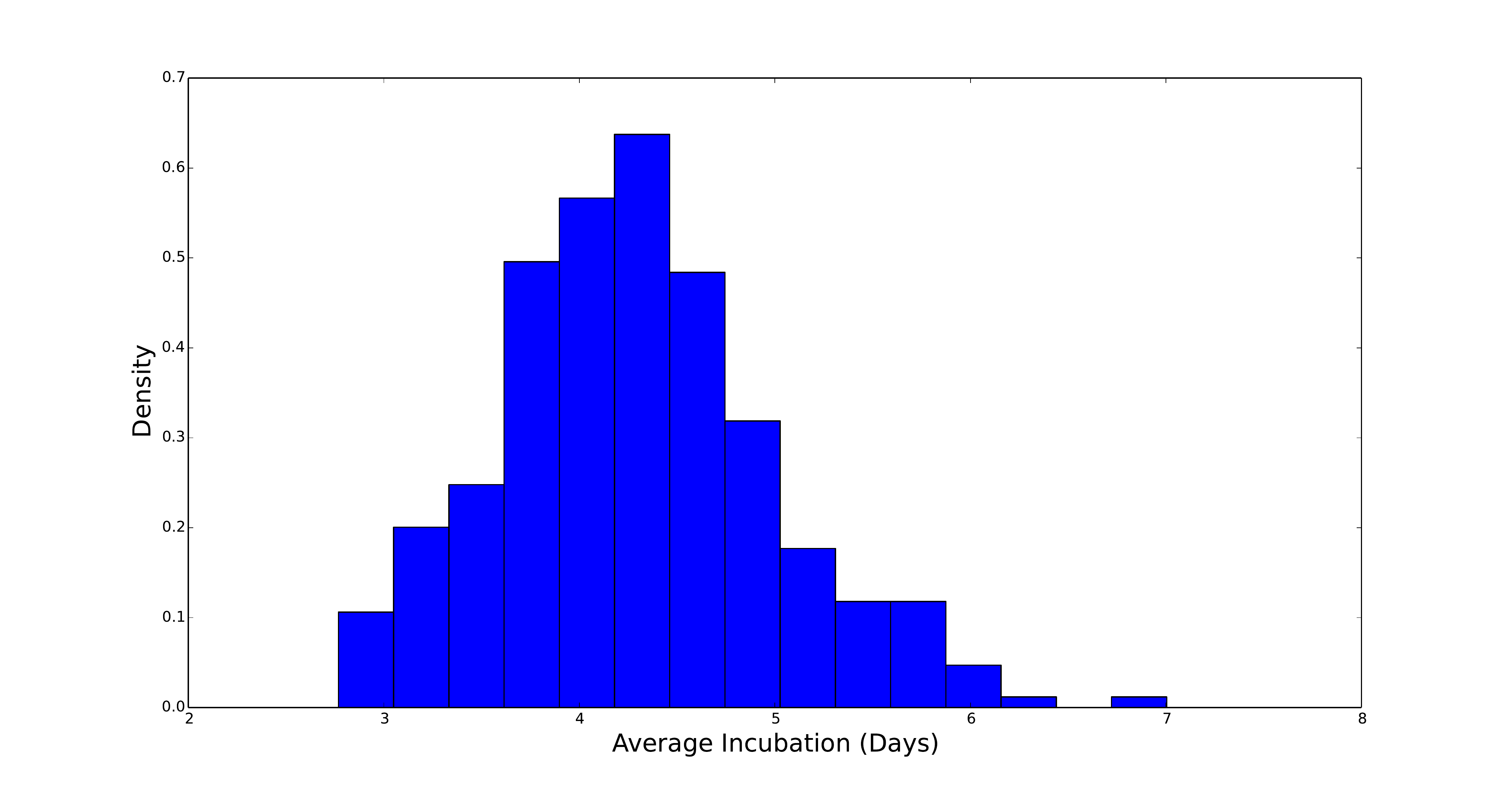} \\
  \includegraphics[scale=0.2]{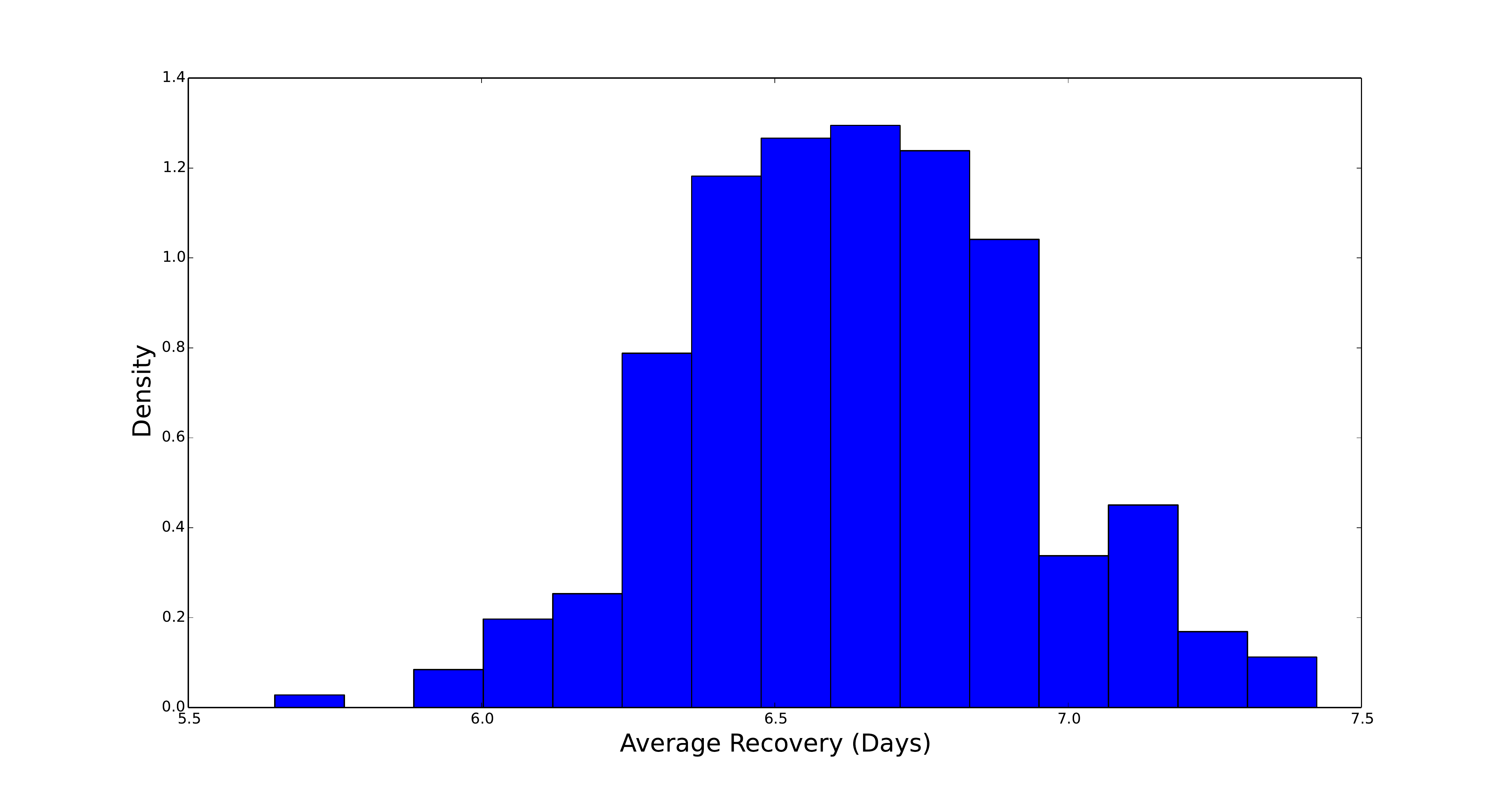} &
  \includegraphics[scale=0.2]{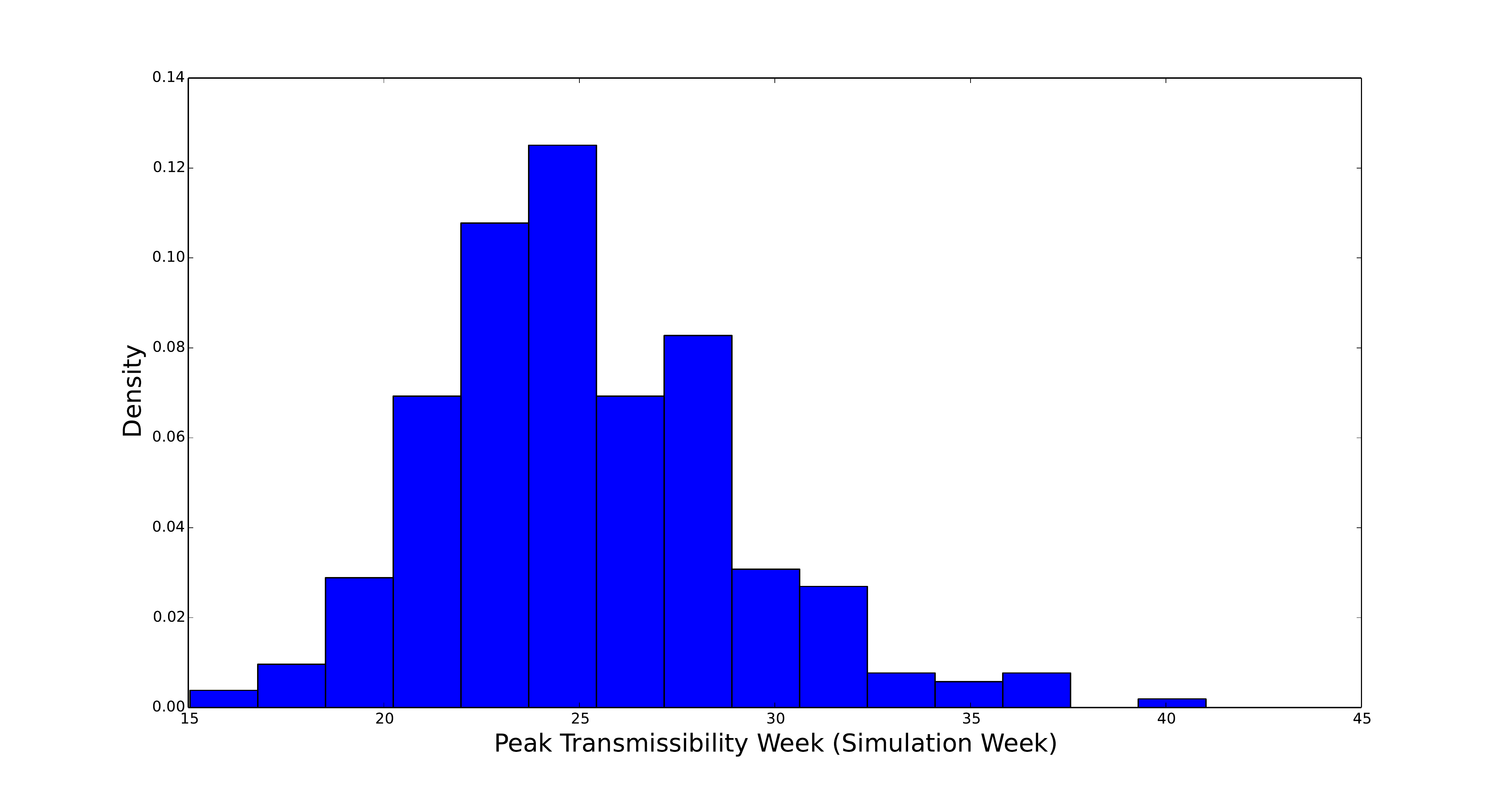} \\
  \includegraphics[scale=0.2]{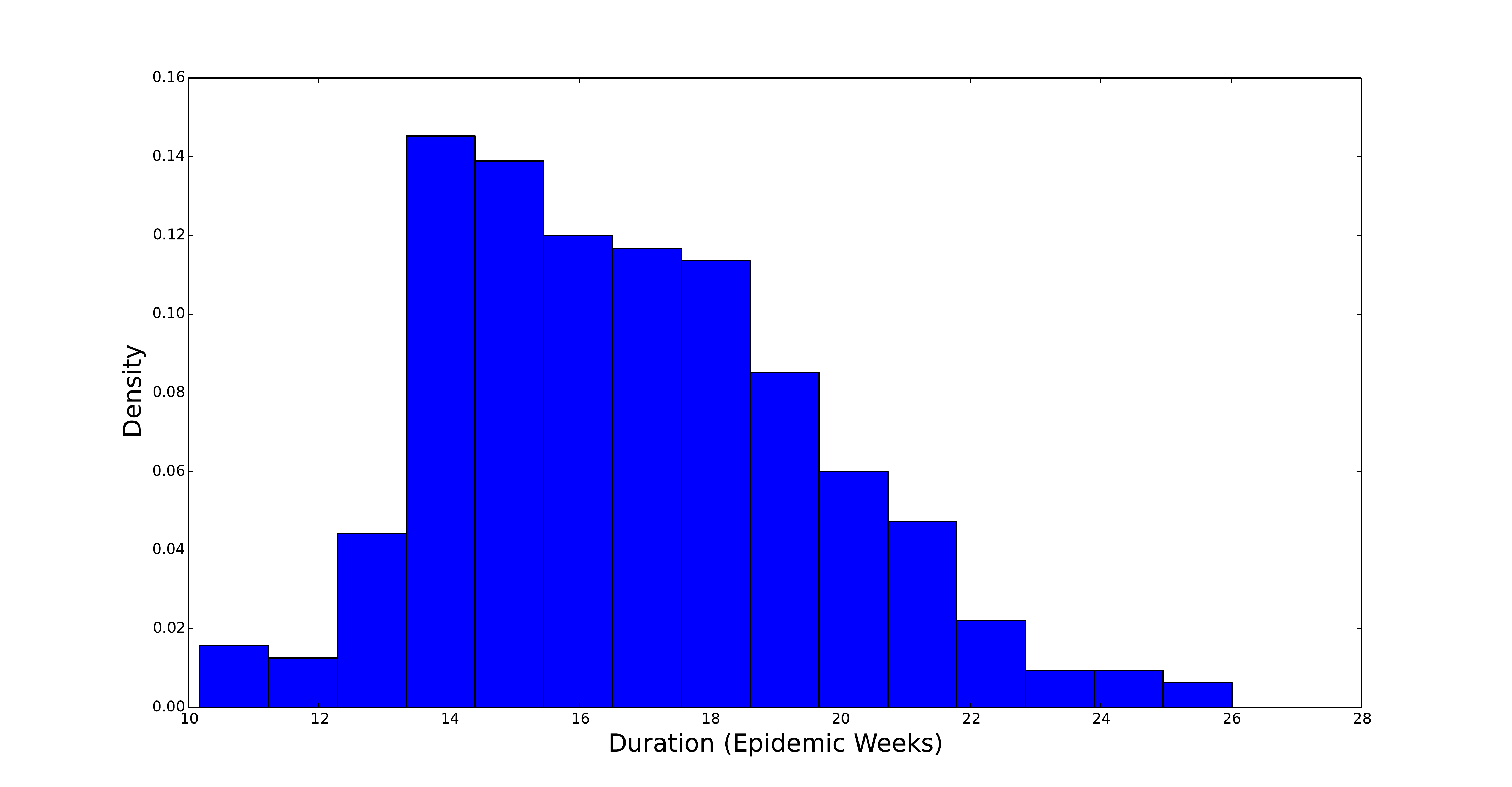} & 
    \end{array}$
    \caption{Histograms of model parameter prior distributions. All histograms
      were generated from 300 samples of $\pi_0(\mathbf{p})$. (TOP LEFT)
      Histogram of the marginal distribution for the average transmission
      time, measured in days. The rate parameter, $\beta_0$, is then the
      inverse of this average time. (TOP RIGHT) marginal distribution for the
      average incubation time, measured in days. The rate parameter in our
      $S^{\nu}\!EIR$ model, $\theta$, is then the inverse of this average
      time. We see that this distribution is concentrated over 3--6 days and
      skewed toward longer incubation times. (MIDDLE LEFT) Histogram of the
      marginal distribution for the average recovery time, measured in days.
      The rate parameter in our $S^{\nu}\!EIR$ model, $\gamma$, is the inverse
      of this average time. We see that this distribution is concentrated over
      6--7 days and skewed toward longer incubation times. The prior
      distribution for $\gamma$ is more concentrated than the distributions
      for $\theta$ and $\beta_0$ which means that the ILI data determine this
      parameter more exactly. (MIDDLE RIGHT) Histogram of the marginal
      distribution for the peak of the transmissibility function, $\beta(t;
      \beta_0, \alpha, c, w)$. The parameter $c$ here is represented in
      \emph{weeks since the beginning of simulation}. Thus, a value $c = 16$
      corresponds to the peak transmissibility during the
      48\textsuperscript{th} epidemiological week. We see that this
      distribution is concentrated over 20--30 weeks into the simulation and
      skewed toward late in the simulation. (BOTTOM LEFT) Histogram of the
      marginal distribution for the duration of heightened transmissibility.
      The parameter $w$ is represented in weeks. A value $w = 14$ corresponds
      to 16 weeks of elevated transmission. We see that this distribution is
      concentrated over 14--20 weeks and skewed toward longer periods of
      elevated transmission.}
    \label{fig:marginal_histograms}
\end{figure}

From the joint prior, $\pi_0(\mathbf{p})$, we can examine samples from the
marginal priors to examine our method's forecast for traits of the
\emph{average} influenza season. In Figure~\ref{fig:marginal_histograms} we
show histograms of samples from a few of these marginal priors. We see that
our methods have determined that the average base time of transmission is 2--5
days, the average incubation time is 3--7 days, and the average recovery time
is 6--8 days. This automatically lets us know that the recovery rate is
tightly specified by our prior whereas the base transmission and incubation
rates are not. Since our $S^{\nu}\!EIR$ model includes a variable transmission
rate, we also include the marginals for the week of peak transmissibility,
$c$, and the duration the transmission rate is elevated, $w$. For our prior,
the duration is centered over 14--20 weeks, while the center is between
simulation week 20 and 30, corresponding to the epidemiological weeks 52 and
10 of 2013 and 2014, respectively.

\begin{figure}[!h]
  \includegraphics[width=\textwidth]{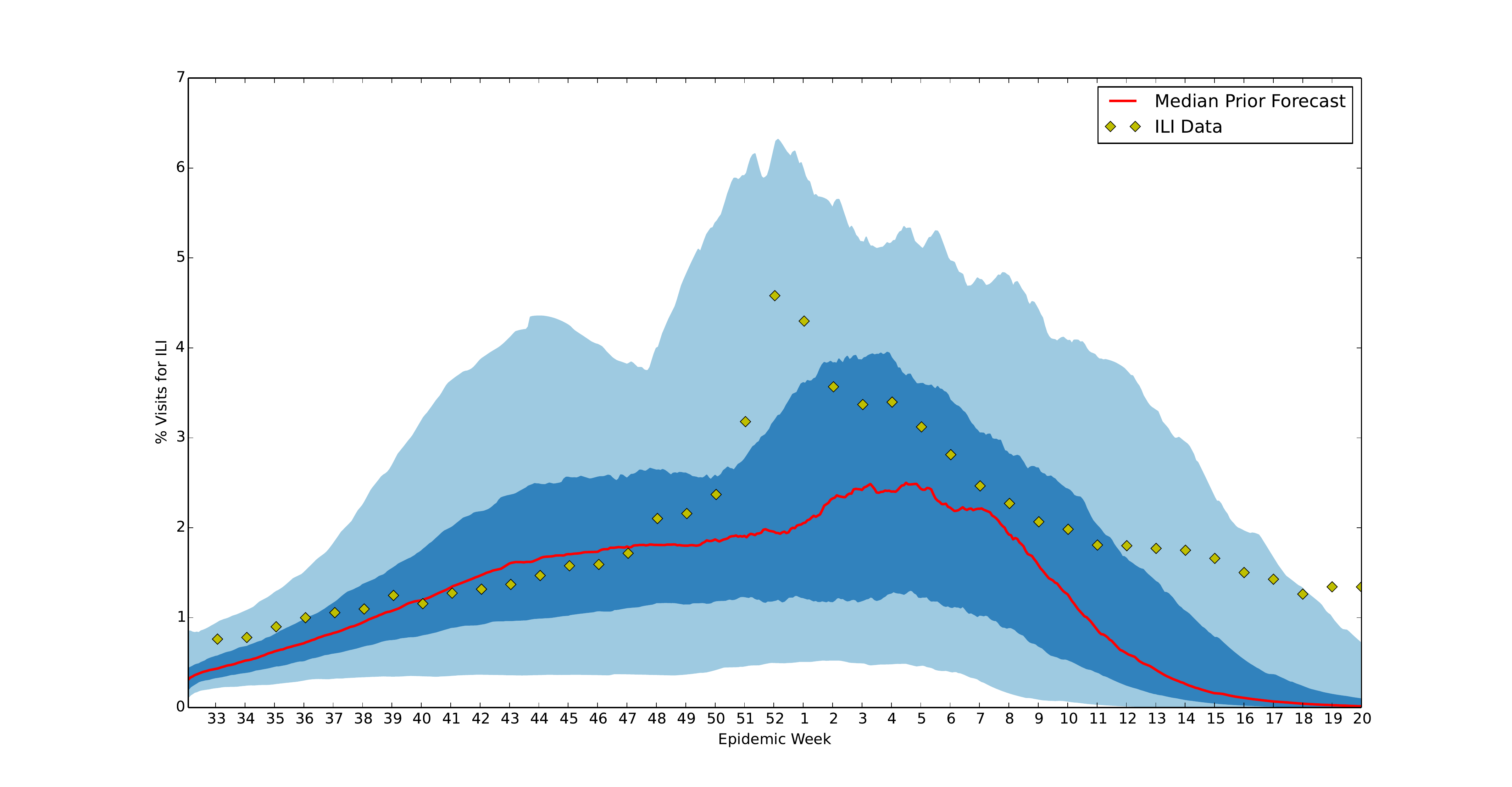}
  \caption{{\bf U.S. ILI prior forecast for 2013--2014 flu season.} This
    figure shows the prior forecast along with the 2013--2014 ILI data. Note
    the potential for an early and late peaking influenza season. The red line
    represents the median forecast from 300 samples of the prior. The dark
    blue and light blue regions represent the 50\% and 90\% credible regions
    centered around this median, respectively. Credible intervals were also
    generated from 300 samples of the prior.}
    \label{fig:US_SEIRprior}
\end{figure}

Sampling 300 parameterization from the log-normal prior $\pi_0(\mathbf{p})$
leads to the prior forecast for the 2013--2014 influenza season shown in
Figure~\ref{fig:US_SEIRprior}. One can notice that our prior forecast allows
for a wide range of peak times and sizes. In general, the earlier the peak the
smaller its forecast height. It is also apparent that our forecast tapers off
quickly after the peak occurs.

\subsection*{Forecast analysis}

\begin{figure}[!h]
  \includegraphics[width=\textwidth]{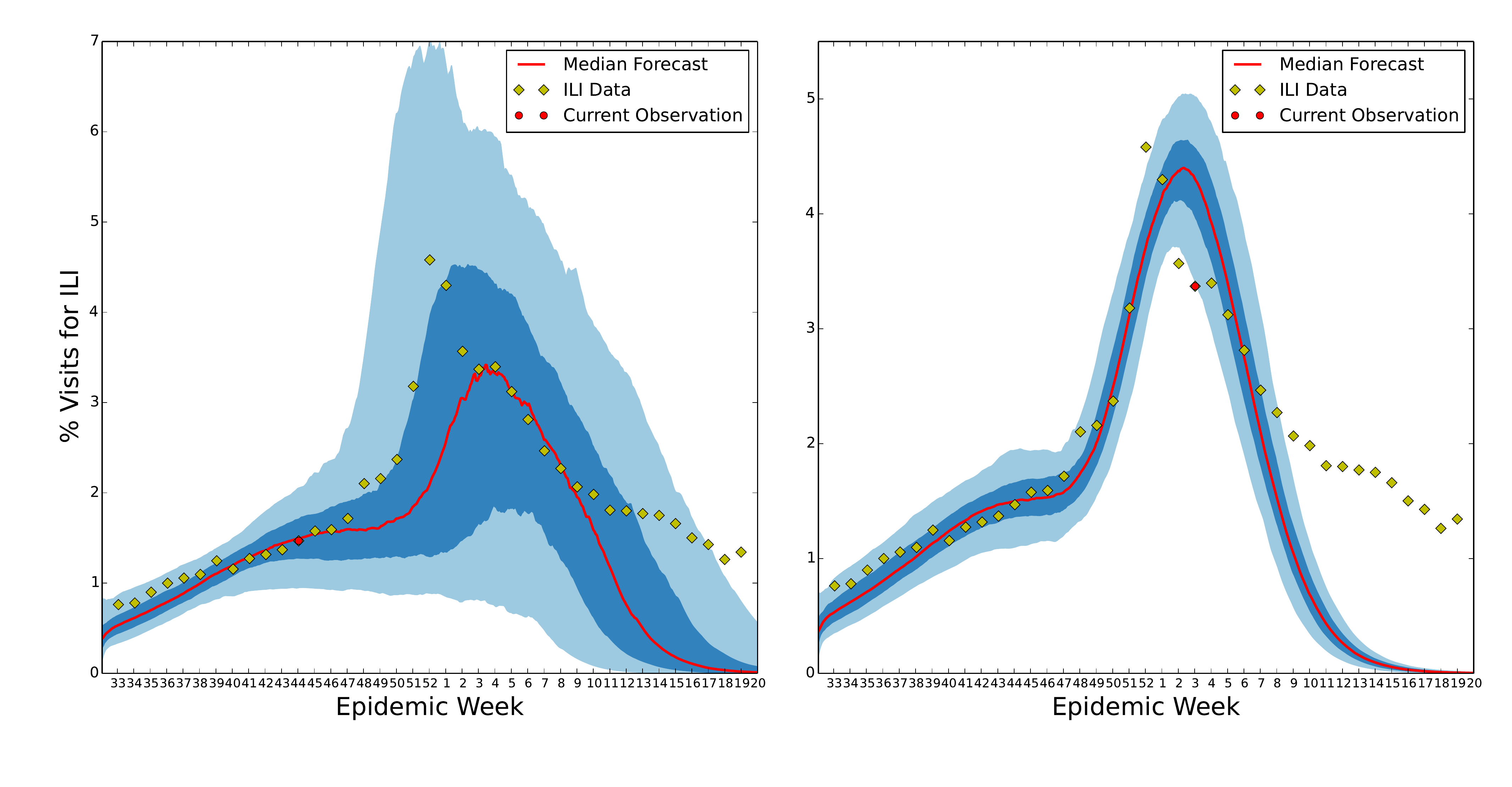}
  \caption{{\bf U.S. ILI forecast for the 2013--2014 flu season made during
      the 43\textsuperscript{rd} (left) and 2\textsuperscript{nd} (right)
      epidemiological weeks.} In each plot the dark blue region represents the region
    centered about the median in which 50\% of forecasts fall, the light blue
    region represents where 90\% of forecasts fall, and the red line
    represents the median forecast. The diamonds represent the 2013--2014 ILI
    data with the current data point marked by a red circle.}
  \label{fig:SEIR_forecasts}
\end{figure}

\begin{figure}[!h]
  \includegraphics[width=\textwidth]{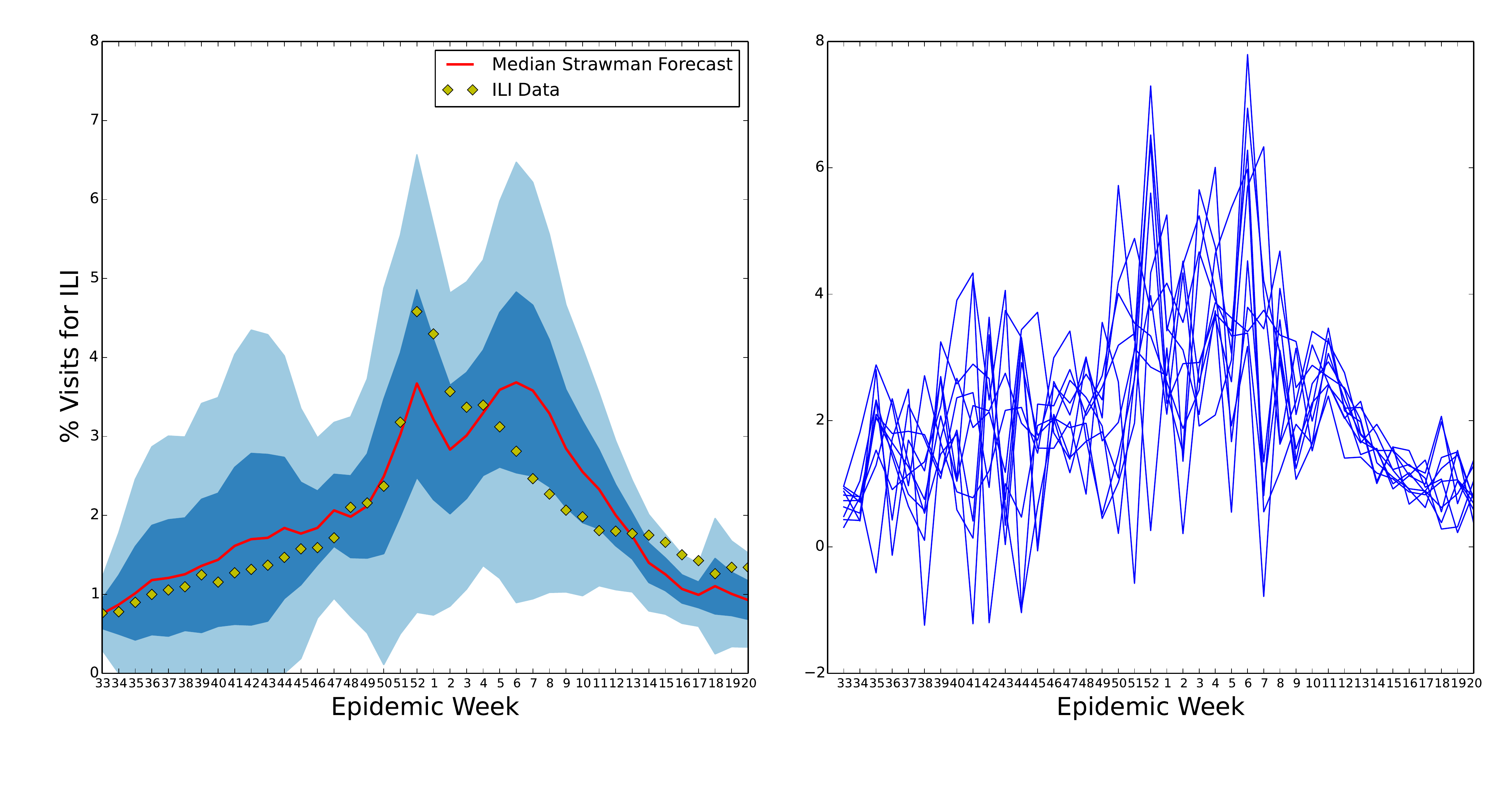}
  \caption{{\bf U.S. ILI Straw Man forecast.} This figure shows the
    results of our straw man prediction based on averaging past flu
    seasons. Since this forecast of the 2013--2014 influenza season
    was made using only the statistics of the sample mean and sample
    standard deviation from previous season's ILI observations it's
    credible intervals (left) do a good job of containing the current
    influenza season. However, this is at the cost of having large
    credible intervals representing a large amount of uncertainty in
    the forecast. Also, this forecast does not include any causal
    model of influenza spread. There is, therefore, no correlation
    between the forecast at successive time points. This is seen when
    sampling time series from this forecast (right).}
  \label{fig:strawman_timeseries}
\end{figure}

In Figure~\ref{fig:SEIR_forecasts} we show the results of our
forecasting process for two different weeks during the 2013--2014
influenza season. Note, the performance of the seasonal $S^{\nu}\!EIR$
model is drastically reduced once the peak of the influenza season has
past. However, before the peak, the model forecasts a range of
possible influenza scenarios that include the 2013--2014 season. In
Figure~\ref{fig:strawman_timeseries}, we show the forecast resulting
from the straw man approach for the 2013--2014 influenza season. In
this section, we analyze both of these forecast's performance using
the measures described above.

\subsubsection*{Quantitative Accuracy}

\begin{figure}[!h]
  \includegraphics[width=\textwidth]{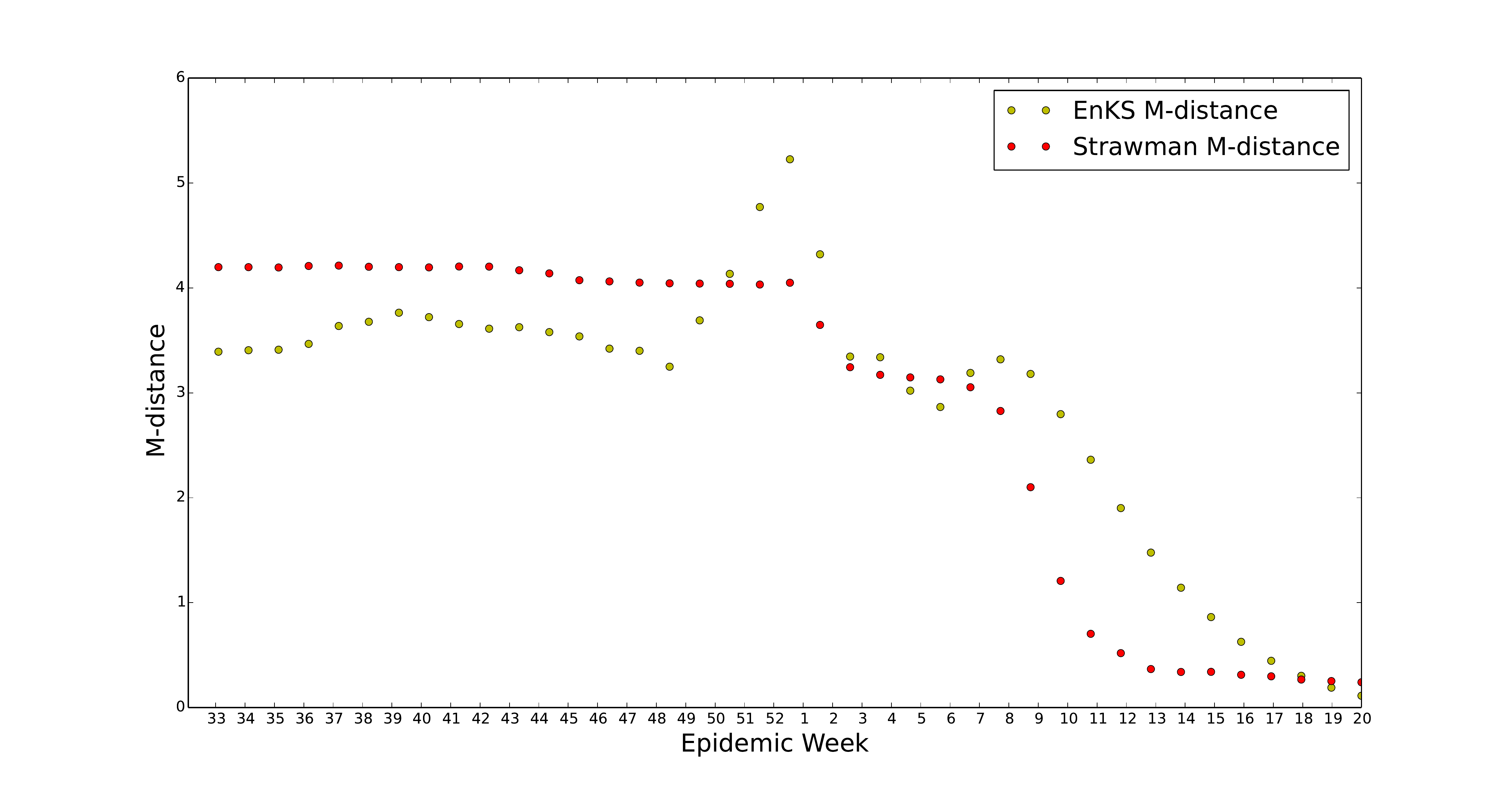}
  \caption{{\bf $S^{\nu}\!EIR$ with enKS vs. Straw Man forecast for the
      2013--2014 U.S. ILI data.} The M-distance between U.S. 2013--2014 ILI
    data and the two forecasts is plotted. The M-distance between the forecast
    and ILI data is calculated for each epidemiological week until the end of the
    influenza season. So the M-distances at week 36 uses the forecast
    observations from week 36 of 2013 to week 20 of 2014 and the ILI data from
    week 36 of 2013 to week 20 of 2014. The M-distances plotted for the
    straw man prediction use sample covariances and means calculated from 300
    time series draws of the straw man forecast. Due to the lack of causal
    relations included in the straw man model this measure of accuracy is
    significantly lower in the early season for the straw man prediction.
    This figure shows that the data assimilation forecast has a noticeably
    smaller M-distance, and therefore is quantitatively better, than the straw
    man model for the early influenza season. Once the influenza season peaks
    the success of the forecast breaks down due to model error. It is
    interesting to note that due to the enKS data assimilation our $S^{\nu}\!EIR$
    forecast seems to attempt self-correction, i.e. the M-distance is
    increasing and then decreases.}
    \label{fig:SEIR_vs_strawman_sample_Mdist}
\end{figure}

In Figure~\ref{fig:SEIR_vs_strawman_sample_Mdist} we show the
successive M-distances computed for our forecast and for 300 samples
from the straw man forecast. We notice that up until the peak of the
2013--2014 influenza season, the data-assimilative forecasting scheme
has a noticeably smaller M-distance than that of the straw man
forecast. However, after the peak of the influenza season, during week
52 for 2013--2014, the straw man shows considerably smaller
M-distance. This is due to the seasonal $S^{\nu}\!EIR$'s inability to
taper off slowly from the peak of the flu season. After the peak, our
model has exhausted its susceptible proportion of the population, and
the infected proportion rapidly goes to zero. A possible point of
confusion in this analysis is the sharp drop off of the M-distance as
the end of our forecast horizon is approached. This is due to the
decreased dimension of the data being forecasted. During week 17, the
forecasts only need to predict the ILI data for 3 more weeks and
distances in this 3-dimensional space grow much slower as a function
of week-by-week error.

\subsubsection*{Qualitative Accuracy}

Each week, given the 300 infected time series from the analyzed
ensemble, we gain 300 samples of the epidemic peak percent infected,
the epidemic start time, the epidemic duration, and the week of the
epidemic peak. From these 300 samples of these quantities of interest,
we estimate the 5\%, 25\%, 50\%, 75\%, and 95\% posterior
quantiles. The start of the influenza season is defined to be the
first week that ILI goes above 2\% and remains elevated for at least 3
consecutive weeks. The end of the influenza season, used to calculate
the duration, is when ILI goes below the 2\% national baseline and
remains there. This gives us a convenient weekly summary of our
influenza forecast with uncertainty quantified by 95\% posterior
credible intervals about the median.

\begin{figure}[!h]
  \includegraphics[width=\textwidth]{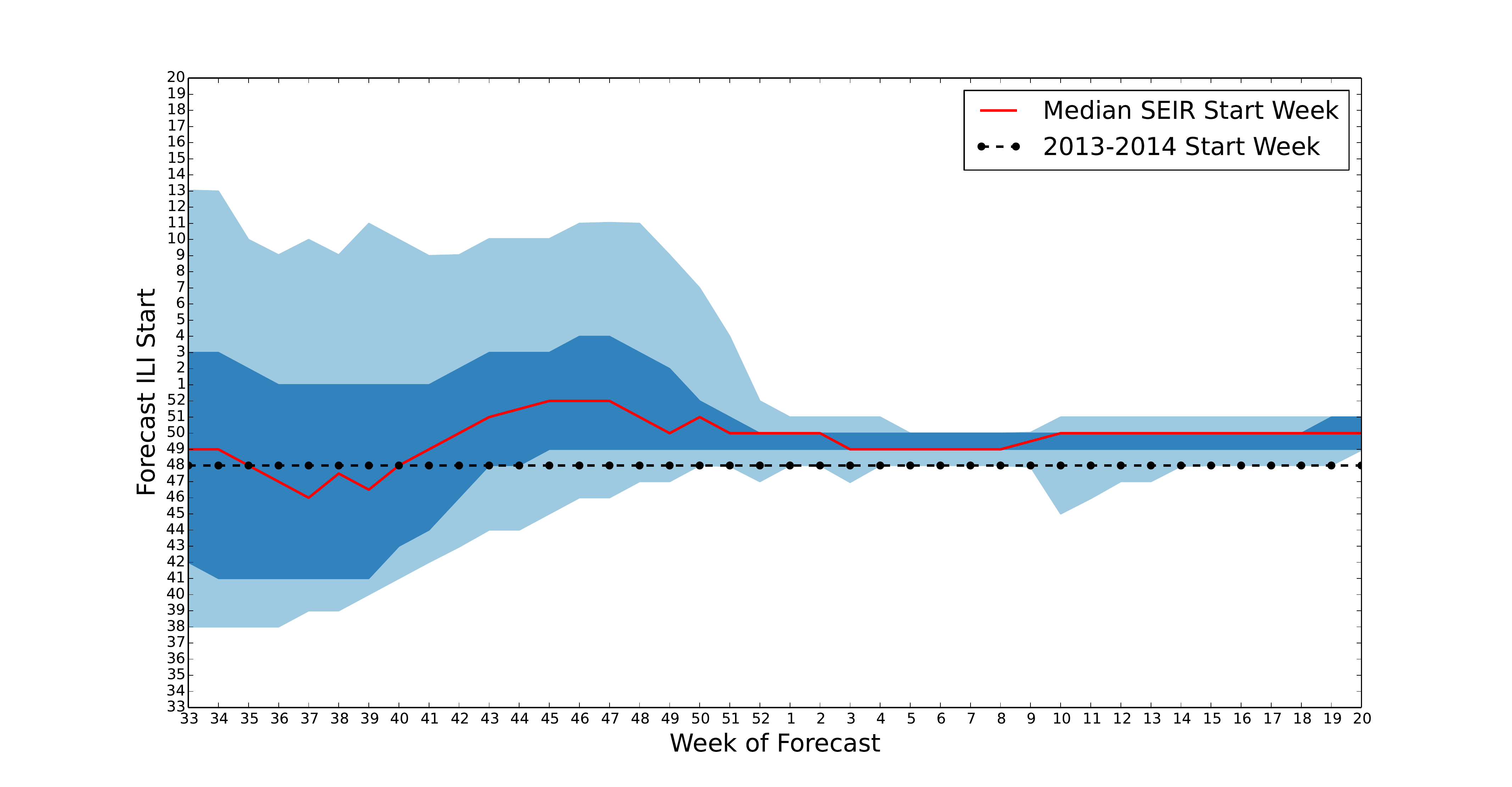}
  \caption{{\bf $S^{\nu}\!EIR$ start week quantiles for 2013--2014 U.S. ILI.}
    50\% and 90\% credible interval estimates of the influenza season start
    week are plotted along with the median. Each week, as new ILI data become
    available the forecast is revised. This causes the uncertainty in our
    forecast to diminish. However, due to the model's inability to maintain an
    elevated ILI level past the peak, we see that late in the flu season, the
    model adjusts by pushing the start week later into the season. This causes
    an overestimation of the start week that worsens as the season
    progresses.}
    \label{fig:enks_QOIseries_stwk}
\end{figure}

A time series plot of the start week credible intervals for our
seasonal $S^{\nu}\!EIR$ forecast is shown in
Figure~\ref{fig:enks_QOIseries_stwk}. A similar plot for the start
week credible intervals forecast using the straw man model would show
a constant distribution with median start week forecast at the
38\textsuperscript{th} epidemiological week. We can see for
Figure~\ref{fig:enks_QOIseries_stwk} that the high probability region
for start week, as forecast by our epidemiological model, is usually
1--2 weeks after the actual 2013--2014 start week. However, the actual
start week is contained within the 95\% confidence region until a week
or two after the peak of flu season. This region is also seen to
constrict as ILI and Wikipedia observations are assimilated.

When compared to the forecast start week from the straw man model, our
$S^{\nu}\!EIR$ model seems like a favorable tool. The straw man's
forecast does not assimilate current observations and thus its
predicted start week is constant.  Two things should be mentioned
about calculating these credible intervals for the straw man
forecast. First, since a sample from the straw man forecast has no
week-to-week correlations, the weekly forecasts can vary greatly from
week to week. This is a problem when computing the start week for the
influenza season since a given straw man sample does not remain above
2\% for consecutive weeks. Second, for similar reasons the duration
cannot even be defined for one time series sample of the straw man
forecast.

\section{Discussion}

\subsection*{Summary of method}

We have outlined an approach to forecasting seasonal influenza that
relies on modern ensemble data assimilation methods for updating a
prior distribution of a disease transmission model. The method used a
dynamic compartmental model of influenza spread that has been used in
previous research
\cite{ross_prevention_1910,stroud_semi-empirical_2006,hyman2003modeling}
but is applicable to any compartmental model of disease with a
regularly updated public health data source.

We evaluated the accuracy of our forecast using the M-distance, based
on the Gaussian likelihood of observations, and the deviation of time
series of quantiles for a set of quantities of interest arising from
flu dynamics. Both of these methods were used on our data assimilation
approach and on a much simpler forecast using estimated normal
distributions. The application of these measures of accuracy, combined
with our specific approach to data assimilation with a dynamic model
of the influenza dynamics allowed us to highlight model inaccuracies
that can then be improved in the future.

Though a statistically simple tool, the inclusion of a straw man
forecast as a baseline to evaluate our data assimilation scheme's
usefulness is indispensable when evaluating measures of
accuracy. Especially for some measure, such as the M-distance, it is
difficult to tell whether or not a value implies the forecast
performed well without a baseline. We hope that the approach of
measuring a forecast against a baseline becomes established practice
in future developments of epidemic forecasting.

The differential equation representation of influenza dynamics,
modeled proportions of the population as susceptible,
exposed/non-infectious, symptomatic/infectious, and
recovered/immune/removed. Thus, the model did not allow for any
re-infection of influenza, which is thought to be biologically
accurate for at least a single strain of flu
\cite{alfaro2013deterministic}. We also modeled the effect of
heterogeneity in the influenza contact network and seasonal variation
in the transmissibility of flu. Our method of data assimilation
adjusted the allowable parameterizations and initializations of this
model as ILI data became available.

The forecast was made up of actual realizations of our $S^{\nu}\!EIR$
model used. This has the arguable advantage of highlighting observed
sections of the ILI data stream that differ drastically from the
model's assumptions. However, since the model state is not adjusted at
each ILI data point directly, the forecast with an incorrect model
eventually diverges from the data.

To iteratively update the prior distributions of parameterizations and
initializations, we used an ensemble Kalman smoother. This was
observed to significantly pull the model toward a subset of
parameterizations and initializations that agreed well with the
data. Since the model seemed to be reasonably adjusted toward
observations and retained a significant amount of ensemble variation
in the forecast, there is strong evidence that the assimilation scheme
works well. The challenge now is to arrive at a model that more
accurately represents influenza dynamics perhaps by including
considerations made in
\cite{alfaro2013deterministic,del_valle_modeling_2013,del_valle_mixing_2007,stroud_spatial_2007,bajardi2011human,lee_computer_2010}.

Our quantitative measure of forecast accuracy is motivated by the
Gaussian likelihood function and has been used, in many instances, to
assign a value to the distance from some predicted distribution with a
fixed mean and covariance. This is exactly the setting we are in, when
we make the Gaussian assumptions inherent in the Kalman filter
methods. The application of the M-distance in this instance showed
that our model performed better than the simple straw man forecast in
the beginning of the season but then systematically diverged from the
late season ILI data.

Besides demonstrating the accuracy of our forecast at capturing
overall dynamics, we also quantified our forecating method's ability
to accurately estimate quantities of interest relating to the impact
of a given influenza season. We showed how the time series of forecast
median and posterior credible intervals for the season's start week,
peak week, duration, and peak level changed over time. This measure in
particular demonstrated the advantages of having an underlying
mechanistic model as compared to the purely statistical normal
approximation forecast as used in the straw man forecast.

\subsection*{Future improvements and lessons learned}

This work shows the viability of using a data assimilation method to
sequentially tune a model of disease dynamics. However, it also
highlights the need to use caution when adjusting the model to match
data. If balances inherent in the model are not maintained during each
adjustment step, it is possible to forecast data accurately with a
model that is incorrect (e.g., one that has no single realization that
will reproduce the data up to data error). The upside of this approach
is that if only the model parameterization and initialization are
adjusted, this type of forecasting process allows one to identify the
assumptions of the model that diverge from observations. This is an
important tool, to advance models to more accurate representations of
reality, that could be ignored if data assimilation methods are used
to adjust a model's state and parameterization throughout the
forecast.

The method proposed here, which maintain $S^{\nu}\!EIR$ balances
during assimilation, are not the only possible methods of maintaining
the population balances assumed in a compartmental disease model. More
research needs to be done on the best way to adjust a model to
observations while maintaining an accurate representation of disease
model balances. Moreover, if the goal is to create forecasts for
multiple seasons, forecasting from initial conditions will not always
be viable. It remains an important open research question as to how
far in the past one should optimally start forecasts from. The farther
in the past a forecast is made from, the more dynamics of the model
and data are assimilated. The downside of this is that considering too
much of the models dynamics can impose unnecessary restrictions on the
prior, leading to a divergent forecast.

A major concern for our epidemiological model is the systematic
divergence from the data at the end of the influenza season. This
divergence is evident in the optimal fitting done with our
$S^{\nu}\!EIR$ model on historical ILI data. The CDC, in collaboration
with World Health Organization and the National Respiratory and
Enteric Virus Surveillance System, also releases some data about flu
strain circulation during the season \cite{_overview_2012}. These
data show that often there are one or two outbreaks of secondary
dominant influenza strains in the late season. We hypothesize that
these secondary strains are a primary cause of the heightened tail in
the ILI data and we will investigate a multi-strain influenza model
\cite{alfaro2013deterministic} in future forecasting work.

\section*{Acknowledgments}

This work was inspired by the CDC's hosting of the 2013--2014
\emph{Predict the influenza season challenge}. We would like to thank
the organizers of this challenge and the participants for the wealth
of insight gained through discussions about disease forecasting and
monitoring.

This work is supported in part by NIH/NIGMS/MIDAS under grant
U01-GM097658-01 and the Defense Threat Reduction Agency
(DTRA). Wikipedia data collected using QUAC: this functionality was
supported by the U.S. department of Energy through the LANL LDRD
Program. LANL is operated by Los Alamos National Security, LLC for the
Department of Energy under contract DE-AC52-06NA25396. The funders had
no role in study design, data collection and analysis, decision to
publish, or preparation of the manuscript. Approved for public
release: LA-UR-14-27259

\bibliographystyle{plain}
\bibliography{CDCflu}

\end{document}